\documentclass[conference]{IEEEtran}
\usepackage{booktabs} 
\usepackage{multirow}
\usepackage{enumerate}
\usepackage{eurosym}
\usepackage{amsmath}
\usepackage{url}
\usepackage{hyperref}
\usepackage{flushend}
\usepackage[numbers]{natbib}

\DeclareMathOperator*{\argmin}{arg\,min}
\DeclareMathOperator*{\argmax}{arg\,max}

\newcommand{\miniskip}{\vspace*{-.5\baselineskip}}

\ifCLASSINFOpdf
  \usepackage[pdftex]{graphicx}
  \graphicspath{{./graphics/}}
  \DeclareGraphicsExtensions{.pdf,.jpeg,.png}
\else
  \usepackage[dvips]{graphicx}
  \graphicspath{{./graphics/}}
  \DeclareGraphicsExtensions{.eps}
\fi

\begin{document}
%
\title{Towards Building a Knowledge Base of Monetary Transactions from a News Collection}

\author{\IEEEauthorblockN{Jan R. Benetka}
\IEEEauthorblockA{Norwegian University of \\Science and Technology\\
benetka@idi.ntnu.no}
\and
\IEEEauthorblockN{Krisztian Balog}
\IEEEauthorblockA{University of Stavanger\\
krisztian.balog@uis.no}
\and
\IEEEauthorblockN{Kjetil N{\o}rv{\aa}g}
\IEEEauthorblockA{Norwegian University of \\Science and Technology\\
kjetil.norvag@idi.ntnu.no}}


%


\maketitle

\begin{abstract}
We address the problem of extracting structured representations of economic events from a large corpus of news articles, using a combination of natural language processing and machine learning techniques.  The developed techniques allow for semi-automatic population of a financial knowledge base, which, in turn, may be used to support a range of data mining and exploration tasks.
The key challenge we face in this domain is that the same event is often reported multiple times, with varying correctness of details.  We address this challenge by first collecting all information pertinent to a given event from the entire corpus, then considering all possible representations of the event, and finally, using a supervised learning method, to rank these representations by the associated confidence scores. A main innovative element of our approach is that it jointly extracts and stores all attributes of the event as a single representation (quintuple). 
Using a purpose-built test set we demonstrate that our supervised learning approach can achieve 25\% improvement in F1-score over baseline methods that consider the earliest, the latest or the most frequent reporting of the event.
\end{abstract}


%
\IEEEpeerreviewmaketitle

\section{Introduction}
\label{sec:introduction}

Financial columns are an essential part of every major news portal. Even people who are not employed in the business sector tend to enjoy catchy headlines about acquisitions of start-ups by the big market players. Information about such economic events is partially captured in a (semi-)structured form, for example, in Wikipedia list pages\footnote{\url{https://en.wikipedia.org/wiki/List_of_mergers_and_acquisitions_by_Alphabet}} and in domain-specific knowledge bases, like CrunchBase. These resources, however, are typically limited to a particular genre of business entities and to a handful of transaction types (e.g., CrunchBase focuses on startups and considers only investments, funding, and acquisitions as financial relations). 
Furthermore, the above resources are constructed manually and thus require a continuous editorial work in order to remain up-to-date.  
News collections contain millions of articles and it is not humanly possible to explore and extract all information about economic events manually. Realistically, only the most prominent transactions with high publicity are likely to be extracted and organized.  Yet, it is the whole space of transactions that provides a complete picture about economic entities.

\begin{figure*}
	\centering
		\includegraphics[width=1\textwidth]{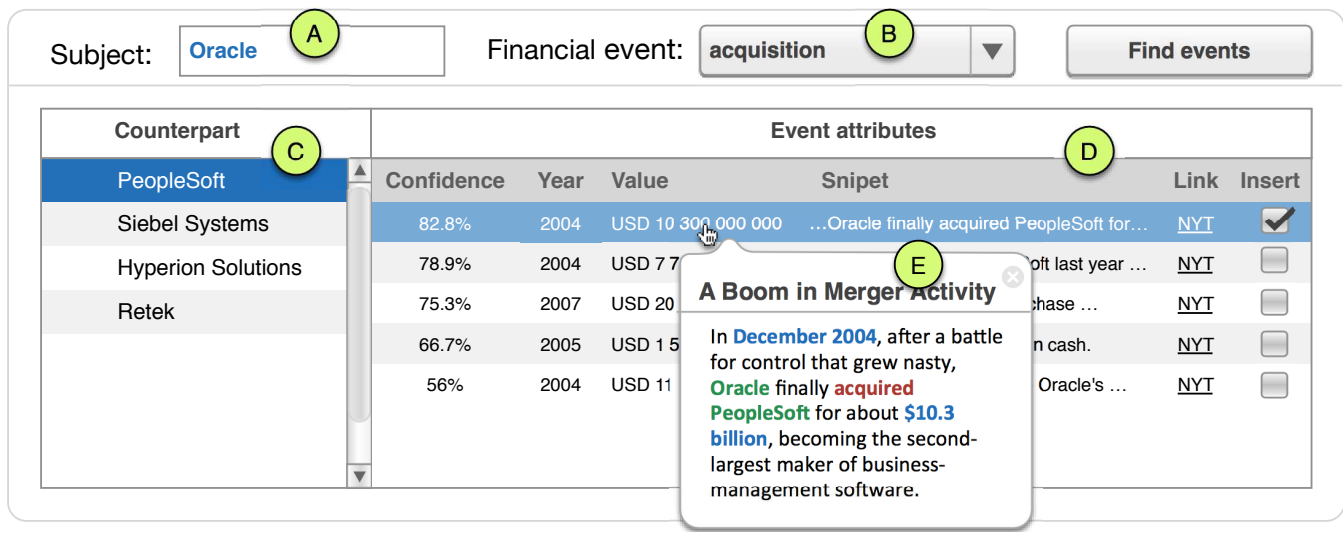}
		\caption{An interactive system for populating a financial knowledge base from a news corpus. For a selected subject entity (A) and financial relation (B), the system finds object entities (C) and lists the extracted events with attributes and an associated confidence score (D); the origins of the extracted data can be checked (E).}
		\label{fig:kbp-tool}
\end{figure*}

Having a comprehensive knowledge base (KB), which organizes information about monetary transactions in a structured and semantically meaningful way would, therefore, be of great value.  This knowledge base could be utilized, among others, for mining and exploring financial entities and trends.  
Our interest is in developing an automated approach that is able to assist in populating a financial knowledge base with previously unseen events, and with updating existing events, if new facts surface. Imagine, for instance, that a person responsible for updating a KB has a tool similar to the one depicted in Fig~\ref{fig:kbp-tool}. By selecting a company (A) and a particular financial relation (B), this tool would mine a predefined set of sources (e.g., The New York Times corpus) and discover companies which are in the given relation with the query company (C). For each of the returned companies, it would identify the attributes of the financial event (D). One challenge KB population faces is due to the very nature of news: they tend to report on the same event multiple times, with slight differences. As we will show later in this paper, simply trusting either the earliest or the latest reporting of the event does not yield the best results. The proposed tool would extract all possible interpretations of the event, as they appear in the text sources and would assign a confidence score to each of these interpretations. This information, together with an interactive preview of the original text (E), would help the editor in deciding which event to include in the KB and which to ignore. Importantly, the editor would update the entire event record with a single click, as opposed to manipulating individual attribute values.

We address the aforementioned challenges by first extracting information from news articles using a natural language processing pipeline.  The proposed pipeline is rather typical in terms of its architecture and components, but is tailored specifically to the financial domain.  It comprises monetary value recognition, economic event recognition, named entity recognition, date extraction, and semantic role labeling steps.

The main conceptual and technical novelty of the paper lies in that all attributes of an event are extracted jointly and a single structured representation is created from them.
We start with grouping all sentences together from the entire corpus that discuss a given event. From these, we generate all possible structured representations of the event, i.e., quintuples comprising subject, predicate, object, monetary value, and date.  All elements of the quintuple are extracted collectively, from a single sentence (which serves as provenance).  To select a single structured representation (quintuple), we employ a supervised learning approach with a set of innovative features to rank the possible quintuples, and then the one with the highest confidence score is chosen. Importantly, our approach can also identify when none of the candidate quintuples would serve as an accurate representation of the event, using a  confidence threshold.
We demonstrate the effectiveness of our method using a purpose-built test collection.

In summary, this work makes the following contributions:
(1) we present a natural language processing pipeline tailored to financial information extraction,
(2) we develop a supervised learning approach and a rich set of features for ranking representations of an economic event and selecting the best one, 
(3) we provide a test dataset and evaluation methodology, and
(4) we perform an experimental evaluation and offer insights on our methods.
All resources developed in this paper are made publicly available at \url{https://github.com/benetka/kbmt}.

\section{Related Work}
\label{sec:related}

The present work lies in the intersection of information extraction, news stream monitoring, financial text mining, and knowledge base population. \\

\textbf{Event extraction} is a specialized branch of information extraction \cite{DBLP:journals/ftdb/Sarawagi08} that has attracted a lot of attention in recent years.
Automated extraction techniques play a crucial role in aiding humans in knowledge-intensive activities in various domains, including global crisis monitoring~\cite{tanev:2008:RTN}, and algorithmic trading~\cite{Nuij:2014:AFI}.  The main approaches and implementations of event extraction from text are well summarized in~\cite{hogenboom:2011:OEE}.
Our work focuses on the extraction of financial transactions.  This is not an entirely unexplored research area. 
\citet{hogenboom:2013:SIE} present a semantic-based pipeline for the detection of economic events (SPEED).  They utilize a traditional language processing pipeline combined with an ontology of economic concepts extracted from Yahoo! Finance.  SPEED is focused solely on the information extraction part; unlike our approach, the authors do not deal with the ambiguity introduced by multiple and possibly conflicting mentions related to one event.	
\citet{Vossen:2014:NRH} describe the NewsReader project and the design of a system aiming at representing events in news streams in a knowledge graph. NewsReader aligns extracted information on a timeline in a story-telling fashion, which is convenient for visual browsing of the data. The paper offers overall statistics of extracted data, but no evaluation is performed. Given the chronological nature of news, the temporal dimension is a common perspective for event exploration \cite{Bgel:2015:TWT}. \citet{Strtgen:2012:ESA} combine time and location for deriving events, however, compared to our work, they do not consider multiple reporting of the same event. \\

\begin{figure*}[t]
	\centering
		\includegraphics[width=0.99\textwidth]{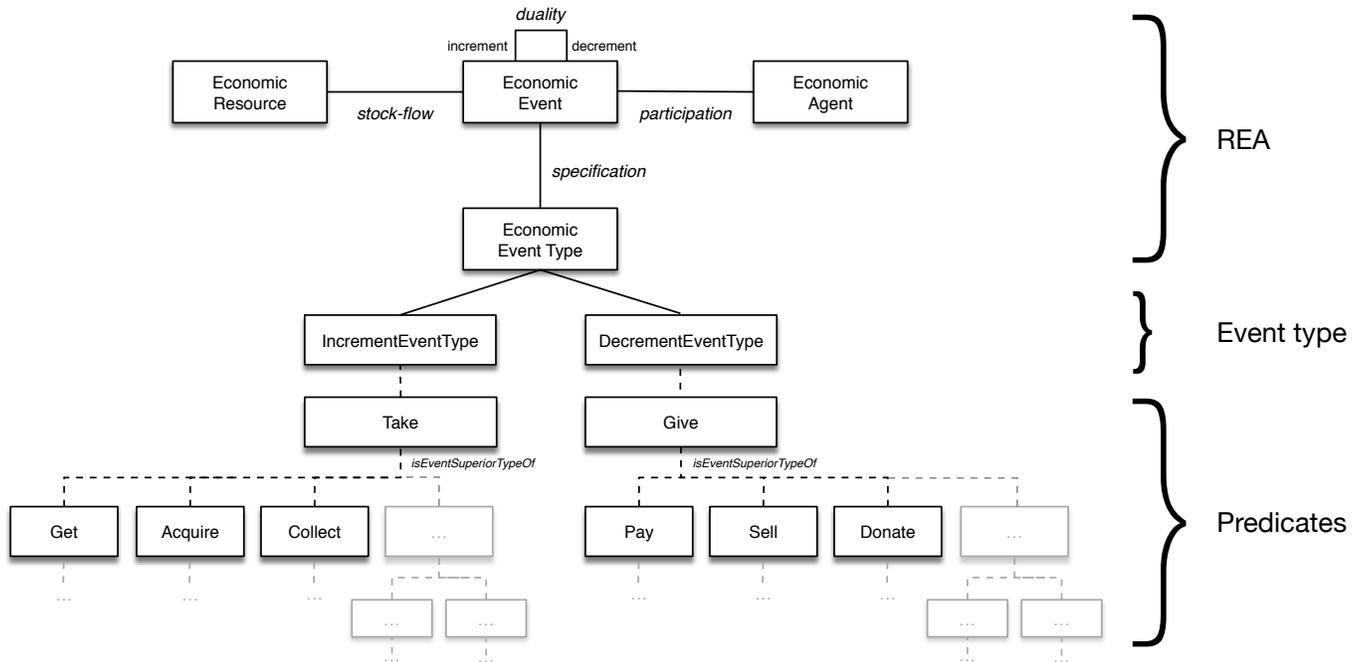}
		\caption{Overview of the Ontology of Economic Events (OEE). Note that the bottom part shows only an excerpt from the instances.}
		\label{fig:ontology}
\end{figure*}

\textbf{Ontologies} are a means to formally model knowledge in the form of (hierarchical) classes of concepts and relations between them.
Domain-specific ontologies can provide fine-grained conceptualization for a specific field of knowledge, e.g., biology~\cite{Ashburner-Gene-2000}, music~\cite{Yves-2007}, or law~\cite{legalonto}.
Concerning the financial domain, several ontologies have been proposed.  The Resource Event Agent (REA) ontology, based on the model developed by~\cite{McCarthy}, represents economic events in an organization from an accounting perspective.  
This model was further analyzed from the ontological perspective using Sowa's conceptual terminology~\cite{Sowa} and is widely used since, either in its core form or in extended versions. 
The Timely Ontologies for Business Relations (TOB) framework~\cite{Zhang:2008:TTO} focuses on business relations and extends the well-known YAGO ontology~\cite{Yago} with a means to represent underspecified time intervals.  This feature allows for temporal relation inference.  A pattern-based approach for financial relation extraction from Wikipedia infoboxes is also presented and evaluated in this work.
Finally, there are ongoing efforts towards standardization of financial reporting.  XBRL (eXtensible Business Reporting Language) is a markup language, with provided taxonomies, that is nowadays widely used by some of the world's largest economies~\cite{XBRL}. \\

\textbf{Knowledge bases}, containing rich semantic knowledge about entities, their properties, and relationships, have become great assets for many applications, including semantic search~\cite{Meij:2014:ELR} and business intelligence~\cite{Weikum:2011:ERB}.
Knowledge base construction and maintenance have been of increasing interest in both academia and industry, see, e.g.,~\cite{Carlson:2010:TAN,Yago2,ji:2011:kbp,Balog:2013:MCA,Balog:2013:CCR}.
General-purpose knowledge bases, such as DBpedia~\cite{DBpedia}, Freebase, or YAGO~\cite{Yago}, cover thousands of business entities; however, they contain limited information regarding financial transactions.  
CrunchBase\footnote{\url{https://www.crunchbase.com/}} is a publicly accessible knowledge base containing comprehensive information about startup companies. At the time of writing, it contains about $650$K profiles of people and companies. Information about financial transactions, which are also part of the data set, are restricted to investments, funding, and acquisitions.  Neither of the aforementioned knowledge bases contain provenance information; our approach supplements each extracted record with provenance data.\\


\section{Representing Economic Events}
\label{sec:kr}

Before we proceed with the description of our extraction pipeline, we present the ontology we developed for conceptual organization of economic events.

Our starting point is the REA (Resource, Event, Agent) model~\cite{McCarthy} that is often used as a foundational model for describing business-related concepts; it is briefly introduced in Sect.~\ref{sec:kr:rea}.  To be able to capture more fine-grained semantic distinctions about financial transactions, we extend REA with a hierarchy of economic event types in Sect.~\ref{sec:kr:oee}.   

\subsection{REA}
\label{sec:kr:rea}

REA has emerged from a framework for accounting systems to one of the standard models in the business domain.  The main concepts of this model are \emph{resources} (e.g., services or money), \emph{events} (e.g., transactions), and \emph{agents} (e.g., companies or people).  Economic events are processes, where economic resources are changing their owners.  It is assumed that there are always two events in a business activity.  One which increases the value of the agent's resources and another, which, in turn, decreases value of another resource belonging to the agent.

\subsection{Ontology of Economic Events}
\label{sec:kr:oee}

In the scope of this project, we deal with a broad spectrum of economic events (i.e., \emph{predicates}) with fine semantic distinctions (e.g., profit-gross).  At the same time, we aim to organize economic events in a hierarchical manner (e.g., get $\rightarrow$ earn $\rightarrow$ profit-gross); subsequent processes can then choose the granularity with which they want the information to be processed.  Currently, there is no ontology available that would allow for such detailed representation of financial activities.  To fill this gap, we propose the Ontology of Economic Events (OEE), an extension to REA;  see Fig.~\ref{fig:ontology} for a graphical overview.
OEE is created using a semi-supervised method that starts with a set of seed verbs and then expands them using the WordNet lexical ontology~\cite{Miller:1995}.

The main class of our economic events ontology is called \emph{EventType}.  Following Hruby~\cite{hruby_model-driven_2006}, we differentiate between two major economic event types: events increasing and decreasing the value of agent's resources.  These sub-classes are called \emph{IncrementEventType} and \emph{DecrementEventType}, respectively.  
We populate these two classes with predicates that represent specific economic events, organized in a hierarchical fashion, using the following procedure.
\begin{enumerate}
	\item Select a set of \emph{seed verbs} that are frequently used in a finance-related context.  We construct this set by extracting verbs (automatically) from all sentences in our corpus that contain a monetary value; then, we select (manually) the most common verbs as predicates that describe a financial transaction (e.g., buy, sell, invest).
	\item For each seed verb:
	\begin{enumerate}
		\item Create an instance of the verb in the ontology.
		\item Find hypernyms (more general words) of the verb in WordNet; these are added as predicates with a parent-child relation to the verb.
		\item Find adjacent terms (word sharing the same hypernym) of the verb in WordNet; these are also added as predicates and linked to the same parent hypernym by a parent-child relation.
	\end{enumerate}
	\item Manually revise the placement of verbs.
\end{enumerate}
This process has led to a hierarchy of $50$ most common business-related verbs, organized into $5$ levels (see the bottom layer on Fig.~\ref{fig:ontology}).\footnote{We wish to point out that predicates from all levels of the hierarchy may be used, not only the leaf nodes.  Obviously, more specific predicates should be preferred over less specific ones.}\\\\

\noindent
\textbf{Example}
Consider the following financial statement: \emph{Apple acquires Beats for \$3.2 billion.}  This information is represented in OEE as three triples: \\

	\resizebox{0.49\textwidth}{!}{
    \begin{tabular}{lll}	
    (Agent) \texttt{Apple}  & \texttt{participates} &  (Event) \texttt{EventID\_1} \\
    (Event) \texttt{EventID\_1} &  \texttt{isClassified} & (IncrementEventType) \\ 
    & & \texttt{acquire}\\
    (Event) \texttt{EventID\_1} & \texttt{inflow} & (Resource) \texttt{Beats} \\
    \end{tabular}
    }\\\\

\noindent
In Sect.~\ref{sec:analysis:ontology} we evaluate the coverage of our ontology using a large news corpus and present further analysis on the usage of predicates in this collection. OEE is made publicly available in OWL format.
\section{Extracting Economic Events}
\label{sec:extr}

\begin{figure*}
	\centering
		\includegraphics[width=0.99\textwidth]{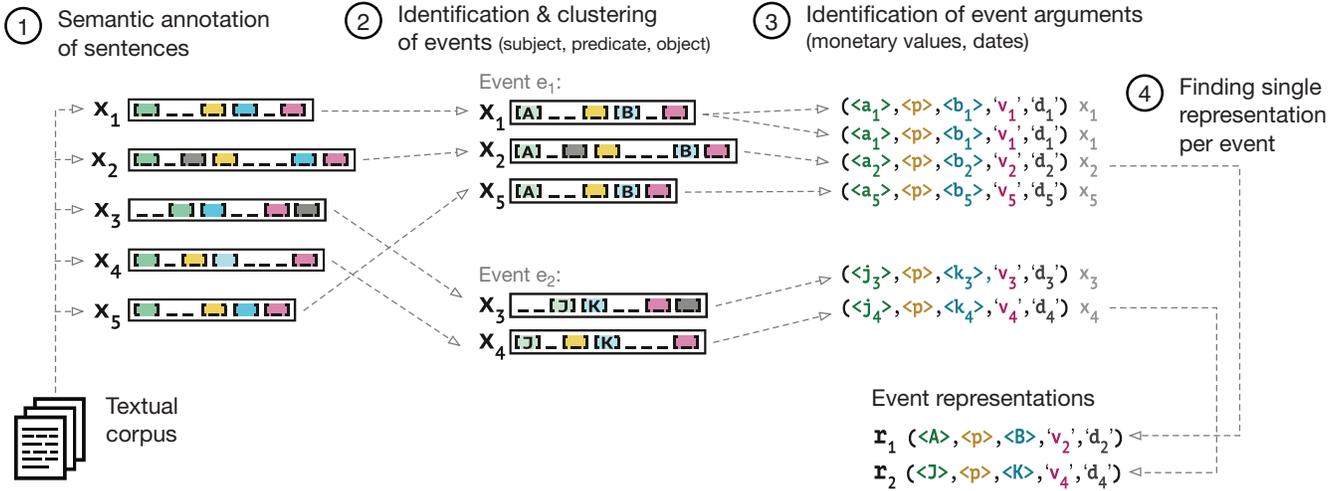}
		\caption{Economic event extraction pipeline.}
		\label{fig:pipeline}
\end{figure*}

Our goal is to extract structured information about economic events from unstructured text (in our case, a large news archive). 
An economic event, as understood in this work, is an unambiguous quintuple:
\begin{equation*}
\label{eq:transaction_quintuple}
\begin{split}
(<subject>, <predicate>, <object>,\\
 monetary\_value, date),
\end{split}
\end{equation*}
where subjects and objects are unique entity identifiers, predicates come from a purpose-built ontology of monetary transactions, and monetary values and dates are normalized literal values.
Our extraction process consists of several steps, organized in a pipeline architecture, as shown in Fig.~\ref{fig:pipeline}.
We deal with the first two steps, semantic annotations (Step 1) and event identification (Step 2), in this section.  Steps 3 and 4 are presented in Sect.~\ref{sec:qtuples}.

\subsection{Semantic Annotations}
\label{sec:extr:annotations}

The first step of our pipeline is responsible for the semantic annotation of text using natural language processing techniques: recognizing financial events, entities, monetary values, and dates.  We operate on the sentence level; sentences serve as provenance information for the extracted information.  Another pragmatic reason for using sentences is that they can be presented as short summaries on the user interface, as it is shown in Fig.~\ref{fig:kbp-tool} (E).
We generate annotations in a sequential order; sentences missing the required piece of information (i.e., monetary value, financial event, or entities) are excluded from subsequent processing steps.  Some of the components in the pipeline have multiple possible configurations; these are evaluated in Sect.~\ref{sec:expeval:trans}.

\begin{enumerate}[(a)]
\itemsep 2pt

\item \textbf{Monetary Value Recognition}\\
Each sentence is tested on the presence of monetary value. 
We define monetary value as a tuple consisting of a numerical value and a currency identifier (e.g., `\euro $1000$' or `two billion US dollars').  A grammar capable of recognition of both verbal and nominal forms of numbers, extended with a list of currency names and symbols, is used for annotation.
Beyond recognition, value and currency normalization are also performed in this step using an extended Numbers Tagger\footnote{\url{https://gate.ac.uk/sale/tao/splitch23.html\#sec:misc-creole:numbers:numbers}} in GATE~\cite{cunningham:2011:gate}.

\item \textbf{Event Recognition}\\
To identify financial transactions in text, we use predicates from a purpose-built ontology of monetary transactions that we constructed in a semi-supervised manner (see Sect.~\ref{sec:kr}) which starts with a set of seed verbs and then expands them using the WordNet lexical ontology~\cite{Miller:1995}. These predicates are used to create a gazetteer for a predicate tagger that, by default, labels verbs. For each of the predicates, the corresponding semantic frame set from PropBank~\cite{PropBank:Palmer:2005} is extracted.  The frame set contains specifications of arguments, referred to as role sets, for possible meanings of the predicate. The specific meaning of each predicate (i.e., role set) is determined later in our annotation pipeline, in the Semantic Role Labeling step.
Further, we extend our annotator with the possibility of recognizing noun predicates as well (e.g., `acquisition of').  We do so by leveraging the NomBank dataset~\cite{meyers:2004:TNP}; each noun in NomBank, provided it originates from a verb, contains an identifier of its source (verb counterpart) in PropBank.  

\item \textbf{Entity Recognition}\\
After having a monetary value and an economic event identified in the sentence, the next step is to recognize the participants of the financial transaction. Since we are only interested in economic subjects, such as companies and organizations, we use a repository of entities assembled from multiple knowledge bases, specifically, DBpedia, Freebase, and CrunchBase; we refer to Sect.~\ref{exp:expsetup:entities} for details.  We resolve each entity mention to its most common sense, i.e., the entity that is most commonly referred to by that mention.  Despite being a naive way of resolving ambiguity, this technique works well in practice~\cite{ji:2011:kbp}.  We consider two settings: one where we require entities to have descriptions, i.e., their knowledge base entry is more complete, and another where such requirement is not imposed. 
\item \textbf{Date Extraction}\\
Each economic event is timestamped with at least one date.  Absolute (e.g., 6/7/2015) and relative (e.g., `two days ago') temporal expressions within the sentence are annotated and normalized by the Stanford Temporal Tagger (SUTime)~\cite{chang:2013:SUTime}.  Note that the sentence might contain multiple explicit temporal expressions, all of which are recorded.  We also consider the article's publication date (this is always available).

\item \textbf{Semantic Role Labeling}\\
Sentences containing all the necessary ingredients mentioned earlier are good candidates for being able to extract a structured representation of the event from them. To verify that all the components (i.e., money, relation verb/noun, and entities) are mutually related, we employ semantic role labeling (SRL), specifically, the system by~\citet{Bjorkelund2009}. SRL is capable of recognizing predicates and their semantic arguments.  Given, for example, the verb \textit{sell}, it will identify a subject (who is selling), an object (to whom) and possibly several other arguments (price, date, manner, etc.) as well.  This final step decides whether the identified components fit a correct semantic pattern.  It is a configuration setting in our pipeline whether the correct semantic roles for monetary value and date are enforced. 

\end{enumerate}

\subsection{Event Identification}
\label{sec:extr:events}

Events, and especially those that involve popular entities, are often reported multiple times in the news media; see Table~\ref{tab:multiplesentences} for an illustrative example. 
Information covered by individual sentences may be redundant, may be conflicting, or may develop over time.  
The second step of our pipeline is responsible for identifying events and clustering sentences that discuss the same economic event.  

We use subject, predicate, object triples to uniquely identify events: $e=(s,p,o)$.\footnote{We note that this identifier allows for a single economic event, with the given predicate, between the two companies; this is currently not an issue in our dataset. It could be easily generalized by including the date as well in the event identifier.}
Subject and object are unique identifiers from the entity repository.  Our entity repository is constructed in such way that the mapping from surface forms of the recognized entities to unique identifiers is unambiguous (see~Sect.~\ref{exp:expsetup:entities}). 
We consider predicates equivalent as long as they have a common ancestor on the second level of the hierarchy; we use the same conditions in our experimental evaluation (cf.~Sect.~\ref{sec:expsetup:methodology}).

\begin{table}[t!]
   \caption{Economic event (Google acquires YouTube) expressed by multiple sentences.}
   \label{tab:multiplesentences}
   \begin{center}
       \begin{tabular}{@{~}p{1.5cm} p{6.5cm}@{~}}
            	\toprule
            	\textbf{Publ. date} & \textbf{Sentence} \\
            	\midrule
			2006-10-11 & Before Google agreed to buy YouTube for \$1.65 billion in stock, it paid \$1 billion for 5\% of AOL... \\					
			2007-02-08 & Google bought YouTube in October for \$1.65 billion. \\ 
			2007-04-05 & YouTube was purchased by Google in November for \$1.6 billion. \\
            	\bottomrule
       \end{tabular}
   \end{center}
\end{table}

\section{Creating Structured Representations of Economic Events}
\label{sec:qtuples}

To this point, we have recognized economic events in sentences, along with their participants (object and subject) and attributes (monetary value and date).  Further, we have grouped sentences together that correspond to the same event.  
What is left for us to do is to create, for each event, a quintuple representing that event.  
This might appear a straightforward exercise at first sight; however, there might be multiple sentences describing the same event (see Table~\ref{tab:multiplesentences} for an example).  Matters are further complicated by the fact that even in a single sentence there might be multiple  financial values or dates, leading to multiple possible interpretations.
We approach this problem in two phases. First, we form one or more quintuples representing the event (Sect.~\ref{sec:qtuples:gen}).  Then, in case there are multiple quintuples, we select one that constitutes the best representation of the event (Sect.~\ref{sec:qtuples:find}).  These phases correspond to Steps 3 and 4 in Fig.~\ref{fig:pipeline}, respectively.

\subsection{Generating Candidate Quintuples}
\label{sec:qtuples:gen}

A single sentence might contain multiple financial values and dates.  In such cases, a quintuple is generated for each possible combination of attributes.  
Formally, let $e$ be an event and $S_e$ the set of annotated sentences describing this event.  Each sentence $x \in S_e$ has the following information extracted: subject ($s$), predicate ($p$), object ($o$), publication date ($d_x$), explicit date mentions ($D_x$), and monetary values ($V_x$).  Note that $s$, $p$, and $o$ are the same across all sentences in $S_e$, because of how sentence grouping works (cf.~Sect.~\ref{sec:extr:events}).  Further note that $D_x$ might be an empty set, while $V_x$ always has at least one element.   
Let then $R_e$ denote the set of possible structured representations for event $e$:
\begin{equation}
   R_e = \big\{ (s,p,o,v,d) | x \in S_e, v \in V_x, d \in D_x \cup \{d_x\} \big\} \nonumber
\end{equation}
For sentences without an explicit date mention, the publication date is used ($d=d_x$); for sentences with one or more dates extracted from the content ($|D_x| \geq 1$), the article's publication date is ignored ($d \in D_x$).

\subsection{Selecting a Single Quintuple}
\label{sec:qtuples:find}

At the end of the processing pipeline, each event $e$ may be represented by a single quintuple.  For events with multiple possible representations (i.e., where $|R_e| > 1$) we need a mechanism to select the quintuple $r \in R_e$ that best describes the given economic transaction.  We present three baseline methods and a supervised learning approach.

\begin{enumerate}[(a)]
	\itemsep 2pt

	\item \textbf{First reporting of the event}\\
	Our first baseline selects the first reporting of event $e$: 
	\begin{equation}
		r^* = \argmin_{d_x \in S_e} \big\{ (s,p,o,v,d) | v \in V_x, d \in D_x \cup \{d_x\} \big\} \label{eq:firstreporting}
	\end{equation}
	In case there are multiple financial values and dates present in the sentence with the earliest publication date, they are chosen arbitrarily.

	\item \textbf{Last reporting of the event}\\
	One might argue that the most recent report is likely to be the most accurate one. Our second baseline method implements this intuition by considering the last reporting of the event.
	This goes analogously to the previous case, except that we write $\max$ instead of $\min$ in Eq.~\eqref{eq:firstreporting}.
	
	\item \textbf{Most frequent reporting of the event}\\
	Intuitively, an information that is repeated multiple times has a strong potential to hold true. The third baseline selects the most frequent reporting of the event. In the case of a tie, the earliest reporting is selected from the pool of the most frequent reportings.
	
	\item \textbf{Supervised learning approach}\\
	We cast the selection of the best quintuple as a regression task and use a machine learning approach.  
	Specifically, we use the Random Forests algorithm~\cite{Breiman:2001:RF}, given its robustness and good empirical performance across a wide range of application domains.
	Our training data comprises a set of instances, $\mathcal{L} = \{ (\mathbf{r}_i, y_i) \}$, where $\mathbf{r}_i$ is a feature vector and $y_i = \{0,1\}$ is a ground truth label corresponding to the quintuple $r_i$.  The learned model is then used to make a prediction $\hat{y}=\varphi(\mathbf{r})$ on an unseen instance $\mathbf{r}$.  We select the quintuple with the highest estimated score:
	\begin{equation}
		r^* = \argmax_{r \in R_e} \varphi(\mathbf{r}) \nonumber
	\end{equation}
	Further, we introduce a confidence threshold $\gamma$, and return the quintuple $r^*$ iff $\varphi(\mathbf{r}) \geq \gamma$.  This is an important feature of our approach, as it allows events to be ignored if there is a lack of support. 
	Moreover, this parameter can be used to control the performance trade-off between precision and recall to suit specific applications.
	The value of $\gamma$ is determined empirically and is set to $0.3$.
	Our feature vector contains a total of $18$ features, developed specifically for this task; it includes 
	(i) simple descriptive statistics (sentence and article length),
	(ii) linguistic features (predicate tense, noun/verb predicate),
	(iii) semantic features, related to automatic as well as explicit semantic annotations (entity identification, semantic roles, temporal value, article category), and
	(iv) cross-document features considering global predicate frequency and attributes across all sentences describing the event (dates and values).
	See Table~\ref{tab:results:feature_desc} for a detailed list.
		
\end{enumerate}

\section{Experimental Setup}
\label{sec:expsetup}

The problem we address in this paper is the automatic extraction of economic events from unstructured text.
This is a restricted and specialized information extraction task for which no standard evaluation resources exist to date. 
Next, we describe the text corpus that serves as our input data (Sect.~\ref{sec:expsetup:textcorpus}), 
the entity repository (Sect.~\ref{exp:expsetup:entities}),
the test collection (Sect.~\ref{sec:expsetup:testdata}),
and our evaluation methodology (Sect.~\ref{sec:expsetup:methodology}).

\subsection{Text Corpus}
\label{sec:expsetup:textcorpus}

We use the New York Times Annotated Corpus (NYTC)\footnote{\url{https://catalog.ldc.upenn.edu/LDC2008T19}} as our input text collection.  This data set contains over $1.8$M news articles spanning over 20 years, beginning in 1987.  Apart from its volume, a great benefit of the corpus lies in the annotations, both automatically generated and manually assigned, accompanying a subset of the articles.  In the scope of this work, we leverage the following annotations as features in our supervised learning step (see~Sect.~\ref{sec:qtuples:find}): publication date, (online) descriptors, and word count.  
We parsed all documents of the NYTC which yielded $2.1$M sentences containing a monetary value.  Out of these, $383$K sentences describe an economic event.  

\subsection{Entity Repository}
\label{exp:expsetup:entities}

We employ an entity repository that is constructed from three sources: DBpedia, Freebase, and CrunchBase.
From DBpedia and Freebase, we only include entities that are of type organization.  CrunchBase contains only companies (over $160$K), so we consider all of them. Some organization names can be expressed in many ways which makes their identification a non-trivial task (e.g. 'The Times', 'The New York Times', 'NYT').  The above-mentioned knowledge bases, however, only hold the official organization names and their unique identifiers. Therefore, on top of the known surface forms, we generate additional name variants using a set of heuristics, similar to those described in~\cite{ananthanarayanan:2008:RBS}.  Finally, we group URIs as well as surface forms together that refer to the same entity and assign a unique identifier to each entity.
Our entity repository contains $989$K unique entities, $1.35$M unique surface forms, and same-as links to $1.24$M DBpedia, Freebase, and CrunchBase URIs in total. An example entry is shown in Table~\ref{tab:entityrepo}.

\begin{table}
   \caption{Entry for the company Skype from our entity repository.}
   \label{tab:entityrepo}
   \begin{center}
   \resizebox{0.49\textwidth}{!}{   
       \begin{tabular}{ll}
            	\toprule
                   \textbf{ID}: & Skype \\
                   \textbf{Surface forms}: & \{Skype, Skype Technologies, Skype Limited\} \\
                   \textbf{URIs}: & \{\texttt{<dbpedia:Skype\_Technologies>}, \\
                   & ~\texttt{<crunchbase:org/skype-technologies>} \\
                   & ~\texttt{<crunchbase:org/skype>}, \\
                   & ~\texttt{<freebase:m/026wfg>},\\
                   & ~\texttt{<freebase:m/06whf7>}\}  \\
            	\bottomrule
       \end{tabular}}
   \end{center}
\end{table}

\subsection{Test Collection}
\label{sec:expsetup:testdata}

We created a test collection by capitalizing on what is already available in CrunchBase.  We hand-picked $30$ target companies from CrunchBase that are known to have participated in financial transactions during the period covered by the NYTC.  
Importantly, the gold standard we need to compare to is not CrunchBase, but what could potentially be extracted from the text corpus (by a human).  Therefore, CrunchBase transactions are checked for their presence in the NYTC; transactions absent from CrunchBase, but covered by the NYTC, are added to the ground truth.
For each target company, we extracted all monetary sentences from the NYTC mentioning the given company and manually grouped sentences by events.  Then, sentences were individually inspected, and a single supporting sentence was selected for each event; the data regarded as ground truth is extracted from this sentence.
In sum, our test data set contains information about investments and acquisitions for $30$ companies, $132$ events in total.

\subsection{Evaluation Methodology}
\label{sec:expsetup:methodology}

We evaluate event extraction as a binary classification task, using standard measures: precision (P), recall (R), and the F1-measure (F1).
The first part of the evaluation (Sect.~\ref{sec:expeval:trans}) is focused on the correct identification of economic events. In order to label an instance as correct, both participating entities as well as the relation type need to correspond with the ground truth. 
When comparing predicates, we considered them equivalent if they had a common ancestor on the second level of the hierarchy.
The second stage of the evaluation (Sect.~\ref{sec:expeval:attr}) examines the extraction of the attributes of economic events. We consider two settings: (1) \emph{strict}, where the financial value and event date have to match exactly, and 
(2) \emph{relaxed}, where a certain tolerance is allowed, specifically, only the year part of date is considered and $10$\% difference in monetary values is allowed.

\section{Evaluating Event Extraction}
\label{sec:expeval}

Evaluation is divided into two main steps:
(1) \emph{event identification}, which considers the extent to which subject-predicate-object triples are successfully identified (Sect.~\ref{sec:expeval:trans}), and
(2) \emph{event extraction}, which focuses on the end-to-end task of creating structured representations of economic events, including their attributes (Sect.~\ref{sec:expeval:attr}).

\subsection{Economic Event Identification}
\label{sec:expeval:trans}

We compare different configurations of our NLP pipeline (in Sect.~\ref{sec:extr}).  
Specifically, we have control over the following options: 
(1) whether noun predicates are also included for event recognition (Y) or only verbs are used (N);
(2) whether semantic roles are enforced for monetary value and date (Y) or not (N);
(3) whether entities are required to have descriptions (Y) or not (N).	

Table~\ref{tab:results:trans} presents the results.  Due to space constraints we do not include all possible combinations, but on having all options `off' or `on' (rows 1 vs. 5), and all but one option `on' (rows 2--4).  We observe that both the addition of noun predicates (NP) and the relaxed treatment of semantic roles (SRL) increase the number of extracted quintuples and events while reducing precision.  Accepting only entities with description (ED) has the exact opposite effect.  The combination of all three methods ensures results with the highest possible recall, without sacrificing precision too much.  We need high recall in downstream processing (where unwanted quintuples can still be filtered out), therefore we use the setting with all options on (row 5) in the remainder of the section.

\begin{table}[t]
   \caption{Economic event extraction results. Highest scores are in boldface.}
   \label{tab:results:trans}
   \begin{center}	
   \begin{tabular}{cccc@{~~}cccc}
	\toprule
	NP & SRL & ED & \#events & \#quintuples & P & R & F1 \\
	\midrule
	N & N & N & 170 & 268 & 0.26 & 0.39 & 0.31 \\ 
	Y & N & N & 185 & 312 & 0.24 & 0.40 & 0.30 \\ 
	N & Y & N & \textbf{316} & \textbf{496} & 0.16 & \textbf{0.44} & 0.23 \\ 
	N & N & Y & 117 & 194 & \textbf{0.37} & 0.38 & \textbf{0.38} \\ 
	Y & Y & Y 	& 217 & 377 & 0.23 & \textbf{0.44} & 0.31 \\ 
	\bottomrule
   \end{tabular}
   \end{center}
\end{table}
		
\subsection{Economic Event Extraction}
\label{sec:expeval:attr}

The second step of the evaluation focuses on the extraction of event attributes, i.e., financial value and date. We shall point out that the extraction mechanism is the same for all methods, but they differ in how a single structured description for the event (quintuple) is selected (cf.~Sect.~\ref{sec:qtuples:find}).
The baseline methods always choose the earliest/latest information.  The supervised method attempts to learn how to select the best quintuple from annotated data; it can also choose not to return any quintuple for a given event.  We use leave-one-out cross-validation, i.e., use all but one company for training and test on the remaining one; this is repeated for all companies in the test set.
 
Table~\ref{tab:results:attr} reports the results for extracting events only (columns 2--4) and extracting attributes as well, using both strict and relaxed evaluation (columns 5--10).  
Focusing on the event extraction part first, we can observe the effectiveness of the filtering mechanism of the supervised learning approach; it doubles precision and improves F1-score by $26$\% (the baselines correspond to the last row in Table~\ref{tab:results:trans}).
Next, when event attributes are also considered, we find again that the supervised learning approach achieves better results in terms of F1-score than any of the baselines. The improvements over the earlier baseline (the better of the two) are $10$\% in strict mode and $25$\% in relaxed mode.

\begin{table}[t]
   \caption{Attribute extraction results for the baselines (BL) vs. our supervised learning approach. Highest scores are in boldface.}
   \label{tab:results:attr}
   \begin{center}
   \begin{tabular}{@{}l@{~}c@{~}c@{~}c@{~}l@{~}c@{~}c@{~}c@{~}l@{~}c@{~}c@{~}c@{~}@{}}
	\toprule
	\multirow{2}{*}{Method}
		& \multicolumn{3}{c}{Events only} 
		&& \multicolumn{3}{c}{Attr. strict} 
		&& \multicolumn{3}{c}{Attr. relaxed} \\
		 \cline{2-4} \cline{6-8} \cline{10-12} 
		& P & R & F1
		&& P & R & F1		
		&& P & R & F1 \\
	\midrule
	BL/earliest 
		& 0.23 & \textbf{0.44} & 0.31
		&& 0.18 & \textbf{0.34} & 0.23
		&& 0.22 & \textbf{0.42} & 0.29 \\
	BL/frequent
		& 0.23 & \textbf{0.44} & 0.31
		&& 0.16 & 0.31 & 0.21
		&& 0.21 & 0.41 & 0.28 \\	
	BL/latest 
		& 0.23 & \textbf{0.44} & 0.31
		&& 0.16 & 0.31 & 0.21
		&& 0.21 & 0.40 & 0.27 \\	
	Our approach
		& \textbf{0.51} & 0.31 & \textbf{0.39} 
		&& \textbf{0.34} & 0.20 & \textbf{0.25}
		&& \textbf{0.49} & 0.29 & \textbf{0.36} \\
	\bottomrule
   \end{tabular}
   
   \end{center}
\end{table}

\section{Analysis}
\label{sec:analysis}

This section provides further analysis of the data and of the results.
Specifically, we check the coverage of our ontology and the frequency of predicates,
measure the importance of individual features, and take a closer look at some successes and failures.

\subsection{Ontology}
\label{sec:analysis:ontology}

\begin{figure}[h!]
	\centering
		\includegraphics[width=0.47\textwidth]{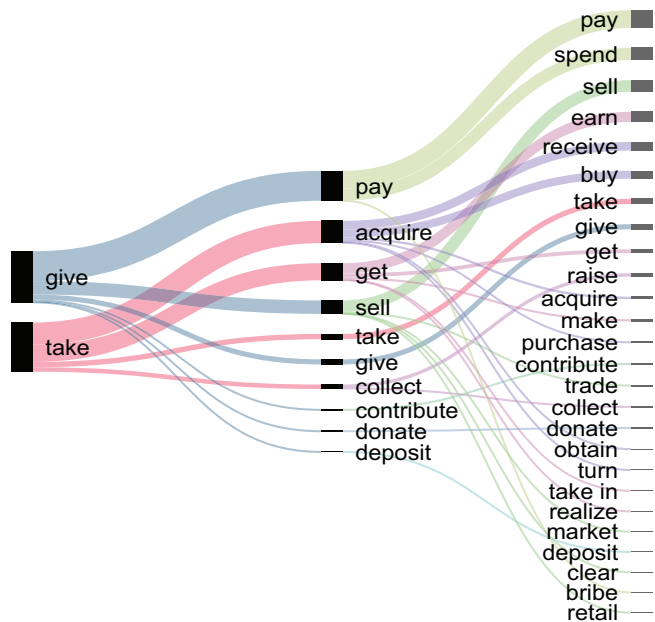}
		\caption{Predicate frequency in the NYTC.}
		\label{fig:pred_frequency}
\end{figure}

The type of each economic event is defined by a predicate in the OEE ontology.  In order to evaluate the coverage of the ontology, we created a list of the most frequent verbs from the $2.1$M sentences of the NYTC with monetary value, manually inspected the top $200$ verbs and deemed $81$ of them as finance-related.  $84$\% of these finance-related verbs is covered by our ontology.
Further, we measured the frequency of the various predicates in the NYTC.  Figure~\ref{fig:pred_frequency} shows the predicates ordered by number of occurrences up to the first three levels of OEE.  The most frequent second-level predicate, \textit{pay}, is mentioned in over $66$K sentences. The average amount of sentences per transaction type from our ontology is $2,339$.

\begin{table*}[ht]
   \caption{List of our features along with their Gini importance.}
   \miniskip
   \label{tab:results:feature_desc}
   \begin{center}

   \begin{tabular}{lp{103mm}lr}
	\toprule
	Feature & Description & Type & Gini\\
	\midrule
	dates\_count &  Quintuple \# with the same date in $R_e$ & numerical & 0.1859\\
	article\_length & Length of the article & numerical & 0.1447\\			
	sentence\_length & Length of the sentence & numerical & 0.1365\\
	sentence\_order & Sentence's position within in the article & numerical & 0.1291 \\
	values\_ratio & Relative freq. of the given monetary value in $R_e$ & numerical & 0.0883\\		
	correct\_fin\_arg & Fin. value is within the correct semantic arg. & binary & 0.0636\\	
	pred\_frequency & Relative freq. of the predicate in the corpus & numerical  & 0.0507\\	
	predicate\_tense & Tense of the predicate  & categorical & 0.0425 \\
	object\_has\_cb\_uri & Object has a CrunchBase URI  & binary & 0.0405\\	
	object\_has\_dbp\_uri & Object has a DBpedia URI & binary & 0.0237\\	
	nytc\_desc\_bus & Article is classified under ``Business" according to the NYTC taxonomy  & binary & 0.0232 \\		
	has\_event\_date & Temp. expression was found within the sentence  & binary & 0.0207 \\		
	correct\_temp\_arg & Temp. value is within the correct semantic arg.  & binary & 0.0186\\
	object\_has\_fb\_uri & Object has a Freebase URI & binary & 0.0156\\	
	is\_noun\_predicate & Predicate is expressed by a verb or a noun  & binary & 0.0096\\	
	subject\_has\_dbp\_uri & Subject has a DBpedia URI & binary  & 0.0067\\
	subject\_has\_cb\_uri & Subject has a CrunchBase URI  & binary & 0.0000\\
	subject\_has\_fb\_uri & Subject has a Freebase URI & binary & 0.0000\\	
	\bottomrule
   \end{tabular}
   \end{center}
\end{table*}

\begin{table*}[ht]
   \caption{Example of a transaction with multiple quintuples: Oracle acquired PeopleSoft.}
   \miniskip
   \label{tab:results:multiple_quintuples_results}
   \begin{center}
   \begin{tabular}{lllllclc}
	\toprule
	Subject & Predicate & Object & Monetary value & Year & Published & Returned by method & Correct \\
	\midrule
	Oracle & acquire & PeopleSoft & \$7.3 billion & 2003 & 2003-11-25 & baseline, earliest  & N \\ 
	Oracle & acquisition & PeopleSoft & \$7.7 billion & 2004 & 2004-10-26 & -  & N \\ 
	Oracle & acquisition & PeopleSoft & \$7.7 billion & 2004 & 2004-10-26& -  & N \\ 
	Oracle & acquire & PeopleSoft & \$1.3 billion & 2004 & 2005-12-23 & -  & N \\ 
	Oracle & acquire & PeopleSoft & \$7.038 billion & 2004 & 2005-12-23 & -  & N \\ 
	Oracle & acquire & PeopleSoft & \$10.3 billion & 2004 & 2007-03-01 & \textbf{supervised learning} & \textbf{Y} \\ 
	Oracle & acquisition & PeopleSoft & \$10.3 billion & 2005 & 2005-06-30 & -  & N \\ 
	Oracle & purchase & PeopleSoft & \$20 billion & 2007 & 2007-03-21 & baseline, latest & N \\ 
	\bottomrule
   \end{tabular}
   
   \end{center}
\end{table*}
 
\subsection{Features}
\label{sec:analysis:features}

Table~\ref{tab:results:feature_desc} lists our features ordered by their Gini importance.
We find that features that consider information from all quintuples for the given event are especially useful (da\-tes\_count and values\_ratio), and so are global predicate statistics (pred\_frequency).  The most important linguistic feature is whether monetary values stand in the correct semantic argument (correct\_fin\_argu\-ment); semantic roles seem far less crucial for dates (correct\_\-temp\_argument).  Article and sentence length are among the strongest features.

\subsection{Successes and Failures}
\label{sec:analysis:success}

We now take a closer look at cases where our supervised learning approach can really make a difference: events for which multiple structured representations (quintuples) are generated.  Our data set contains $24$ such events; the number of quintuples for these range from $2$ to $17$.  The results for these events, using relaxed evaluation, are as follows: the earliest baseline fails in $4$ cases, the latest baseline fails in $5$ cases, while the supervised learning approach was incorrect only in a single case. Table~\ref{tab:results:multiple_quintuples_results} shows a specific example, where the same event is reported multiple times. The supervised learning method was able to identify the correct quintuple.

\section{Conclusions}
\label{sec:conclusions}

In this paper, we have addressed the task of extracting economic events from a large news corpus.  
We have presented a natural language processing pipeline for the semantic annotation of text.  To create a single structured representation for each economic event, we have employed a supervised learning approach and have developed a set of innovative features.  Using a purpose-built test collection, we have demonstrated that our approach is superior to two intuitive baselines, i.e., earliest and latest published information, and can achieve $25$\% improvement in F1-score.

Our work represents an important step towards building domain-specific knowledge bases in an automated manner.  Even if the system may not have reached the necessary level of performance yet for fully automated operation, it could aid human editors in their tasks by displaying verifiable structured records in a ranked order.
The next direction for future work is to consider multiple textual sources, not just a single newspaper.



\bibliographystyle{abbrvnat}
%


\bibliography{jcdl2017-financial}

\end{document}